\documentclass[
reprint,
footinbib,
amsmath,amssymb,
aps,
prd, tikz
]{revtex4-2}

\usepackage{graphicx}
\usepackage{dcolumn}
\usepackage{enumitem}
\usepackage{bm}
\usepackage{float}
\usepackage[many]{tcolorbox}
\usepackage{shuffle}
\usepackage{environ}
\usepackage{physics}
\usepackage{tabularx}
\usepackage{diagbox}
\usepackage{subcaption}
\usepackage{multirow}
\usepackage{hyperref}

\newcommand{\DF}{(DF)^2}
\newcommand{\hp}{\hat{\partial}}
\newcommand{\hco}{\hat{\cal O}}
\newcommand{\tc}{\tilde{c}}
\newcommand{\cT}{{\cal T}}

\font\tenshuffle=shuffle10 \font\sevenshuffle=shuffle7 \font\fiveshuffle=shuffle7 at 5pt
\def\shuffle{{%
    \def\Dshuffle{\mathbin{\hbox{\tenshuffle\char'001}}}%
    \def\Sshuffle{\mathbin{\hbox{\sevenshuffle\char'001}}}%
    \def\SSshuffle{\mathbin{\hbox{\fiveshuffle\char'001}}}%
    \mathchoice{\Dshuffle}{\Dshuffle}{\Sshuffle}{\SSshuffle}}}

\usepackage{CJK}

\begin{document}

\begin{CJK*}{UTF8}{}
\CJKfamily{gbsn}

\title{
Uniqueness and Analytic Structures of Bosonic String Effective Amplitudes
}

\author{Qu Cao(曹趣)$^{1,2}$}
\email{caoqu@westlake.edu.cn}
\author{Fan Zhu(朱凡)$^{3}$}
\email{zhufan25@gscaep.ac.cn}

\affiliation{
$^{1}$Department of Physics,
School of Science, Westlake University, Hangzhou 310030, China \\
$^{2}$Institute of Natural Sciences, Westlake Institute for Advanced Study, Hangzhou 310024, China\\
$^{3}$Graduate School of China Academy of Engineering Physics, No. 10 Xibeiwang East Road, Haidian District, Beijing,
100193, P.R.China}

\begin{abstract}
We revisit the zero-transcendentality sector of bosonic string effective amplitudes with spin-1 external states, conjectured to correspond to a mass-deformed $\DF$ theory, known as the $\DF{+}{\rm YM}$ theory. Imposing gauge invariance, locality, and cyclicity under minimal assumptions uniquely fixes a set of dimension-raising operators and leads to a recursive construction of amplitudes from Yang-Mills amplitudes in the $\alpha'{\to}0$ limit. At finite $\alpha'$, certain derivative operators dressed with gauge invariant and $\alpha'$-dependent factors, what we call {\it inverse operators}, reconstruct the full bosonic string effective amplitudes, yielding compact expressions that universally factorize into tachyon-pole coefficients times Yang-Mills-Scalar amplitudes. This structure holds at arbitrary multiplicity and also extends to the amplitudes of the pure $\DF$, $\DF{+}\phi^{3}$ and $\DF{+}{\rm YM}{+}\phi^{3}$ theories.
\end{abstract}
\maketitle
\end{CJK*}

\section*{Introduction.}
Perturbative string amplitudes are known to exhibit a universal low-energy expansion at tree level~\cite{Huang:2016tag,Stieberger:2009hq}, in which the leading transcendental coefficients coincide for superstring and bosonic string theories at each order in $\alpha'$. While superstring amplitudes reduce to (super) Yang-Mills(YM) amplitudes in the zero-transcendentality sector, bosonic string amplitudes retain nontrivial $\alpha'$ corrections.  The bosonic string effective amplitudes (or simply bosonic amplitudes) in zero-transcendentality sector has been conjectured to correspond to a massive deformation of the $\DF$ theory, dubbed $\DF{+}{\rm YM}$~\cite{Azevedo:2018dgo} theory.

The uniqueness of scattering amplitudes has been widely explored in Yang-Mills theory, gravity, and scalar effective field theories~\cite{Arkani-Hamed:2016rak,Rodina:2016jyz,Rodina:2016mbk,Rodina:2018pcb}, as well as for string world-sheet integrals such as the Veneziano and Virasoro-Shapiro amplitudes~\cite{Veneziano:1968yb,Virasoro:1969me,Shapiro:1970gy} from the bootstrap string program\cite{Caron-Huot:2016icg,Huang:2020nqy,Guerrieri:2021ivu,Geiser:2022icl,Geiser:2022exp,Cheung:2022mkw,Cheung:2023adk,Cheung:2023uwn,Haring:2023zwu,Albert:2024yap,Cheung:2024uhn,Cheung:2024obl,Cheung:2025tbr,Wan:2026pjq}. In contrast, the uniqueness of effective field theories for gauge theories remains largely unexplored. In this work, we address this gap by studying the effective field theory-- $\DF{+}{\rm YM}$ theory-- from the low-energy expansion of bosonic string amplitudes.

Let us begin by reviewing the progress on tree-level string amplitudes. The pure-spinor formalism~\cite{Berkovits:2000fe,Berkovits:2004px,Berkovits:2005bt} leads to the well-known formula~\cite{Mafra:2011nv,Mafra:2011nw,Broedel:2013aza,Broedel:2013tta}, where superstring amplitudes are expanded in a world-sheet integral basis $F$ with coefficients given by (super) Yang-Mills amplitudes,
\begin{equation}\label{eq:purespinor}
    {\cal A}_n^{\rm super}(1, \rho, n{-}1, n)=\sum_{\sigma \in S_{n{-}3}} F_\rho^\sigma A^{\mathrm{YM}}(1, \sigma, n{-}1, n)\,.
\end{equation}
The $A^{\mathrm{YM}}$ carry all of the polarization dependence. The $\alpha'$-dependence, by contrast, is entirely carried by $F_\rho^\sigma=\delta_\rho^\sigma+{\cal O}(\alpha')$.
Bosonic string amplitudes admit an analogous expansion~\cite{Huang:2016tag,Azevedo:2018dgo}, whose coefficients $A^{\DF+\mathrm{YM}}$ are deformed by $\alpha'$ corrections
\begin{equation}\label{eq:bos pure}
    {\cal A}_n^{\rm bos}(1, \rho, n{-}1, n)=\sum_{\sigma \in S_{n{-}3}} F_\rho^\sigma A^{\DF+\mathrm{YM}}(1, \sigma, n{-}1, n)\,.
\end{equation}
The coefficients $A^{\DF+\mathrm{YM}}$ encode an extension of Yang-Mills theory by an infinite tower of higher-derivative operators. It has been pointed out in~\cite{Huang:2016tag, Azevedo:2018dgo} that a particular $\DF{+}{\rm YM}$ Lagrangian~\cite{Johansson:2017srf}, involving pure YM with higher-derivative terms coupled to a colored scalar, generates the polarization dependence of open bosonic
string amplitudes, \textit{i.e.}, the bosonic amplitudes $A^{\DF{+}{\rm YM}}$.
Here, the universality of bosonic string interactions is broken due to the $\alpha'$ dependence of $A^{\DF+\mathrm{YM}}$, which leads to non-transcendental contributions in the stringy corrections.

The rich information encoded in bosonic amplitudes $A^{\DF+\mathrm{YM}}$ has been investigated using field-theoretic methods. The lower-point amplitudes can be computed using the Feynman rules~\cite{Johansson:2017srf}, however, these rules are complicated and tend to obscure the underlying structure of the bosonic amplitudes. Another conventional approach to obtain $A^{\DF+\mathrm{YM}}$ is based on the direct evaluation of world-sheet integrals. For the bosonic string, integration-by-parts (IBP) techniques are required to handle integrands that are not of $d$-log form. Although the IBP algorithm has been developed in~\cite{He:2018pol, He:2019drm}, this method still difficult to uncover the entire analytic structure of $A^{\DF+\mathrm{YM}}$.

Two fundamental questions remain and motivate the present work: (a). How can these bosonic effective amplitudes be understood from the perspective of effective field theory, and are they uniquely determined? (b). How can these bosonic effective amplitudes be systematically constructed, and do they exhibit simplified structures?

In this letter, we revisit this effective field theory amplitudes, try to explore its uniqueness and analytic structures. The two main results of this letter are as follows: (a). Under mild and well-defined assumptions, we establish a restricted notion of uniqueness for $A^{\DF{+}{\rm YM}}$: while uniqueness in a broad sense does not hold, a well-defined uniqueness can be achieved within a narrower framework. Moreover, the uniqueness of $A^{\DF{+}{\rm YM}}$ provides a structural explanation for the previously mysterious relation~\eqref{eq:new eq super}, originally observed in~\cite{Cao:2025lzv}. (b). We present a systematic and convenient construction of $A^{\DF+\mathrm{YM}}$ from Yang-Mills amplitudes via the inverse operators introduced in this work, and reveal the resulting elegant structures of $A^{\DF+\mathrm{YM}}$ .

\begin{figure}[htbp]
    \centering
    \includegraphics{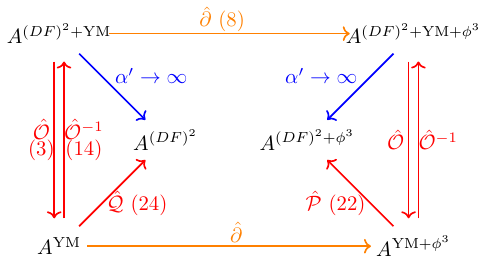}
    \captionsetup{justification=raggedright,singlelinecheck=false}
    \caption{The unified web of tree amplitudes for gauge theories and their mass-deformed and scalar extensions. The operators $\hp$, $\hco$, $\hat{\mathcal{P}}$ and $\hat{\mathcal{Q}}$ are combinations of differential operators with gauge invariant prefactors.}
    \label{fig_gauge_family}
\end{figure}

As an important implication of our results, we also establish a unified web of tree-level amplitudes for gauge theories and their mass-deformed and scalar extensions, as shown in Fig.~\ref{fig_gauge_family}. In the accompanying ancillary \texttt{Mathematica} file, we provide codes for generating all these operators.

\section*{The uniqueness of bosonic amplitudes.}

In recent paper~\cite{Cao:2025lzv}, we find a new formula for superstring amplitudes,
\begin{equation}\label{eq:new eq super}
    {\cal A}_n^{\rm super}=\hco_{[n-1]}{\cal A}_n^{\rm bos}\equiv\left(\sum_{\rho\in [n{-}1]}\prod_{\sigma\in\rho} 2\alpha'\tr_\sigma \hp_\sigma\right){\cal A}_n^{\rm bos} \,,
\end{equation}
which relates the (mixed) bosonic string amplitudes. The sum runs over all products of cycles, denoted by $\rho$, selected from $[n{-}1]{\equiv}\{1,2,\ldots,n{-}1\}$ (so that leg $n$ is fixed), and for the empty cycle we define $2\alpha'\tr_{\varnothing}\hp_{\varnothing}{=}1$. For $|\rho|{>}2$, $\tr_{\rho}$ denotes the trace of field strengths $f_i^{\mu\nu}{:=}p_i^\mu\epsilon_i^\nu-\epsilon_i^\mu p_i^\nu$ along the cycle $\rho$. In the special case $|\rho|{=}2$, we define $\tr_{ij}{=}\frac{1}{2}\Tr(f_if_j)$. The $\hp_\rho$ is differential operator on polarization-dependent coefficients to transmute the gluons into scalars, called transmuted operator~\cite{Cheung:2017ems}, and its definition is given in subsection~\ref{subsec_transop}. The transmutation operators are not all independent: besides cyclic and reflection symmetry, they also obey the Kleiss--Kuijf (KK) relation~\cite{Cheung:2017ems, Kleiss:1988ne} (in the appendix~\ref{app_transop} we give a detailed discussion), \textit{i.e.}, for the $m!$ of operators $\hp_{\sigma(12\ldots m)}$ ($\sigma{\in} S_{m}$), only $(m{-}2)!$ of them are independent. Expanding each $\hp_{\sigma}$ in a Kleiss--Kuijf basis by fixing the minimal and maximal labels of $\sigma$ at the two ends, the ordinary trace prefactors are correspondingly reorganized into right-nested combinations $\tilde{\tr}_\sigma$, and we obtain
\begin{equation}
    \hco_{[n-1]}\equiv\left(\sum_{\tau\in [n{-}1]}\prod_{\sigma\in\tau} 2\alpha'\tilde\tr_\sigma \hp_\sigma\right)\,,
    \label{eq_def_hco}
\end{equation}
where the sum now runs over all products of sequences $\tau$ whose minimal and maximal labels are fixed at the two ends. 
The $\tilde\tr_{12\ldots m}{:=}\tr_{12[3[...[(m-1)m]]]}$ is defined as a right-nested commutator of $\tr_{12\ldots m}$; for example, $$\tilde{\tr}_{12345}{=}\tr_{12[3[45]]}{=}\tr_{12345}{-}\tr_{12354}{-}\tr_{12453}{+}\tr_{12543}.$$

Throughout the paper, a tilde denotes this right-nested commutator projection.

Applying our formula~\eqref{eq:new eq super} to Eqs.\eqref{eq:purespinor} and\eqref{eq:bos pure}, we obtain a differential equation for the field-theory coefficients,
\begin{equation} \label{eq: operator expansion}
A_n^{\rm YM}=\hco_{[n-1]}A_n^{\DF{+}{\rm YM}}.
\end{equation}
This serves as our key starting point for studying the structure of the bosonic amplitudes.

Our first goal is to understand the origin of this equation. Suppose we know nothing about string theory--no vertex operators, no worldsheet integrals-can we still understand why such a relation exists? From our perspective, this question can be viewed within the framework of effective field theory (EFT). The bosonic amplitude corresponds to the non-transcendental sector of the bosonic string effective action~\cite{Huang:2016tag,Azevedo:2018dgo}; it can be obtained by taking the $\alpha'{\to}0$ expansion of the bosonic string amplitude and discarding all transcendental contributions, $\zeta_{i_1,\dots}{\to}0$. Hence, the first step of our analysis is to bootstrap the field-theory relation~\eqref{eq: operator expansion} purely from the EFT point of view.

We consider an EFT amplitude that can be resummed as a function of the inverse string tension $\alpha'$, whose lowest dimensional term corresponds to the Yang-Mills amplitude (we treat $\alpha'$ as dimensionless). Here, we assign the polarization vectors $\epsilon_i$ the same mass dimension as the momenta $p_i$, and measure dimensions in terms of the Mandelstam variables $s_{ij}{=}2p_i\cdot p_j$. Under this convention, the Yang-Mills amplitude has mass dimension two. Then we need to bootstrap the higher mass dimension terms based on the first principle, the gauge invariance and locality. In order to sharp our discussion, we take some hypothesis:
\begin{itemize}
\item  ({\it Smoothness}) The EFT amplitude is a smooth function not only of~$\alpha'$, but also of the kinematic invariants constructed from the polarization and momentum vectors, namely~$\epsilon\cdot p$, $\epsilon \cdot \epsilon$, and~$p\cdot p$.
\item ({\it Bottom-up}) Higher mass dimension terms can be systematically generated from lower mass dimension ones. In particular, the gauge transformations of the lower-dimensional amplitudes play a nontrivial role. Trivial operators such as~$s_{ij}\times$ are physically meaningless, only operators that involve gauge transformation, such as $\epsilon_i\to\epsilon_i+\lambda p_i$, can contribute.
\item ({\it Off-shell current}) The EFT amplitude shares the same off-shell current structure as the Yang-Mills amplitude~\cite{Berends:1987me}, {\it i.e.}, it can be written as the off-shell current contracted with the external polarization~$\epsilon_n$ in the on-shell limit.
\end{itemize}

Besides our three hypotheses, we impose only gauge invariance, locality, and cyclicity on each mass-dimension amplitude. We begin at four points, since at three points gauge invariance allows only the Yang-Mills and $F^3$ amplitudes~\cite{Cheung:2016drk}. For the first nontrivial mass dimension, locality forces the derivatives of lower-dimensional amplitudes to be dressed by gauge-invariant polynomials. Thus the relevant building blocks are transmutation operators~\cite{Cheung:2017ems} multiplied by traces of field strengths, such as $\tr_{12}\hp_{12}$ and $\tr_{123}\hp_{123}$. The four-point ansatz is therefore a linear combination of $\tr_{12}\hp_{12}$, $\tr_{13}\hp_{13}$, $\tr_{23}\hp_{23}$ and $\tr_{123}\hp_{123}$ acting on $A_4^{\rm YM}$; cyclicity fixes all relative coefficients, leaving only the overall normalization later chosen to be $2\alpha'$. At five points the same one-step analysis gives $\sum_{\tau\in[4]}2\alpha'\tilde{\tr}_{\tau}\hp_{\tau}$, while possible double-trace dressings in the one-step ansatz are removed by cyclicity. The new feature at five points is that genuinely higher dimension-raising operators first appear: for example, products such as $\tr_{12}\tr_{34}\hp_{12}\hp_{34}$ raise the mass dimension by two. At the next mass dimension these two-step operators must be combined with the one-step operator acting on the already-constructed mass-dimension-three amplitude; imposing cyclicity again fixes the remaining coefficients and reproduces the product structure in~\eqref{eq_def_hco}.

In summary, the key point here is to construct nontrivial operators that raise the mass dimension. These operators consist of transmutation operators combined with gauge-invariant polynomials. The off-shell current hypothesis allows us to reduce the ansatz to depend only on $(n{-}1)$-particle information for an $n$-point amplitude, while cyclicity ensures the recovery of global symmetry and fixes the remaining undetermined coefficients.

\section*{How to construct bosonic amplitude?}
The most straightforward way to construct the bosonic amplitudes is to invert Eq.~\eqref{eq: operator expansion},
\begin{equation}\label{eq-inverse}
A_n^{\DF{+}{\rm YM}}=\hco^{-1}_{[n-1]}A_n^{\rm YM}\,,
\end{equation}
where $\hco_{[n-1]}^{-1}\hco_{[n-1]}{=}1$. In this section, we first introduce the necessary properties of the transmutation operators and then present a systematic construction of $\hco^{-1}_{[n{-}1]}$, which directly generates the bosonic amplitudes.

\subsection{Transmutation operators.}\label{subsec_transop}
As studied in~\cite{Cheung:2017ems}, they found a set of Lorentz invariant differential operators called {\it transmutation operators} which transmute physical tree-level scattering amplitudes into new ones. The general transmutation operators are organized as two kinds of derivatives
\begin{equation}
    \left\{
    \begin{matrix}
        \text{trace operator: }&\hspace{-5em}\cT_{ij}=\partial_{\epsilon_i\cdot\epsilon_j}\,,\\[5pt]
        \text{insertion operator: }&\cT_{lmn}=\partial_{p_n\cdot \epsilon_m}-\partial_{p_l\cdot\epsilon_m}\,.
    \end{matrix}
    \right.\label{eq_def_calT}
\end{equation}
The trace operator $\cT_{ij}$ reduces the spin of particles $i$ and $j$ by one unit and places them within a new color trace structure. The insertion operator $\cT_{lmn}$ reduces the spin of particle $m$ by one unit and inserts it between two existing bi-adjoint scalars $l$ and $n$ within a color trace structure\footnote{Note that our definition of insertion operator is differ a signature form the one in~\cite{Cheung:2017ems}.}. The general transmutation operators that reduce the spin of multi-particles within a color single-trace structure is defined as
\begin{equation}
    \hp_{\rho_1\ldots\rho_m}^{\sigma(\rho)}{=}\left(\prod_{i=0}^{m-3}\cT_{\rho_{\sigma(m-i)-1}\rho_{\sigma(m-i)}\rho_{\sigma(m-i)+1}}\right)\cT_{\rho_{\sigma(1)}\rho_{\sigma(2)}}\,,
    \label{eq_def_transop}
\end{equation}
where $\sigma{\in} S_m$, the sequence in the subscript denotes the color trace structure, and we call the sequence in the superscript as adjoint orderings of transmutation operators which represent the sequence of reducing the spin of particles. For examples, 
\begin{equation}
    \begin{aligned}
        &\hp_{1234}^{1342}A^{\rm YM}_{12345}=\cT_{123}\cT_{341}\cT_{13}A^{\rm YM}_{12345}=\cT_{123}\cT_{341}A^{\rm YMS}_{13|12345}\\
        &=\cT_{123}A^{\rm YMS}_{134|12345}=A^{\rm YMS}_{1234|12345}\,,
    \end{aligned}
    \label{eq_exp_hp}
\end{equation}
where $A^{\rm YMS}_{1234|12345}$ represents the Yang-Mills-Scalar(YMS) amplitude with $1,2,3,4$ are bi-adjoint scalars~\cite{Cachazo:2013iea}, the second order of scalars is 1234. The situations for multi-trace scalars are similar.

The transmutation operators $\hp_\rho^{\sigma(\rho)}$ with different $\sigma$ should generate the same result while they act on an Lorentz invariant amplitude. In the appendix~\ref{app_transop}, we introduce more properties of transmutation operators. In the entire article, we treat transmutation operators only differ from adjoint orderings as the same, and omit the adjoint orderings.

\subsection{Reconstruction for inverse operators}
\begin{figure}[htbp]
\centering
\includegraphics[]{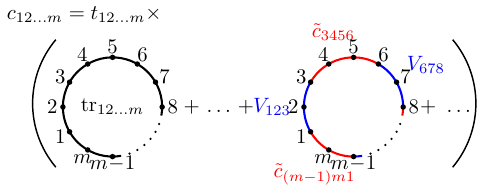}
\captionsetup{justification=raggedright,singlelinecheck=false}
\caption{Recursion definition of $c_{12\ldots m}$. The black circle represents the longest single-trace term $\tr_{12\ldots m}$. The blue and red arcs with nodes $\rho_1\rho_2\ldots\rho_m$ denote $V_{\rho_1\rho_2\ldots\rho_m}$ and $\tc_{\rho_1\rho_2\ldots\rho_m}$, respectively. The colored circle denotes the product of all arcs it contains. The ellipses indicate the set of circles constructed from red and blue arcs, with the requirement that each red arc carries fewer than $m$ nodes.}
\label{fig_crho}
\end{figure}

Apparently, the inverse operator of $\hco$ can be written as an expansion of $1/\hco$ around infinitesimally small $\alpha'$. For example, we have
\begin{equation}
\begin{aligned}
    &\hco_{\{1,2\}}^{-1}=\frac{1}{1+2\alpha'\tr_{12}\hat{\partial}_{12}}
    =1+\sum_{i=1}^\infty(-2\alpha'\tr_{12}\hat{\partial}_{12})^i=\\
    &1+\sum_{i=0}^\infty(\alpha's_{12})^i\left(-2\alpha'\tr_{12}\hat{\partial}_{12}\right)
    =1+\frac{-2\alpha'\tr_{12}}{1-\alpha's_{12}}\hat{\partial}_{12}\,,
\end{aligned}
\end{equation}
where in the second line we have used $\hat{\partial}_{12}\tr_{12}{=}-s_{12}/2$. In the general case, each term in the expansion of $1/\hco$ takes the form
\begin{equation}
    \big(-2\alpha'\tr_{\rho_1}\hat{\partial}_{\rho_1}\big)^{x_1}
    \big(-2\alpha'\tr_{\rho_2}\hat{\partial}_{\rho_2}\big)^{x_2}
    \cdots
    \big(-2\alpha'\tr_{\rho_m}\hat{\partial}_{\rho_m}\big)^{x_m}\,,
    \label{eq_exp1/o}
\end{equation}
where $\rho_i$ and $x_i$ denote arbitrary sequences and integers, respectively. We find that~\eqref{eq_exp1/o} can always be reduced to the form
\begin{equation}
    {\rm coeff}\,\times\,\hat{\partial}_{\sigma_1}\hat{\partial}_{\sigma_2}\cdots\hat{\partial}_{\sigma_l}\,,
    \qquad \forall~i\neq j,\ \sigma_i\cap\sigma_j=\varnothing\,,
    \label{eq_form1}
\end{equation}
where ``coeff'' is a polynomial in $\alpha'$, $\tr$, and $V$, with $V$ defined as a contraction structure of momenta and field strengths
\begin{equation}
    V_{\rho_1\rho_2\ldots\rho_m}:=-p_{\rho_1}\cdot f_{\rho_2}\cdot\ldots\cdot f_{\rho_{m-1}}\cdot p_{\rho_m}\,.
    \label{eq_defV}
\end{equation}
In particular, for $m{=}1,2$, we define $V_{\rho_1}{=}0$ and $V_{\rho_1\rho_2}{=}-p_{\rho_1}\cdot p_{\rho_2}$, respectively. An inductive argument supporting this conclusion is presented in Appendix~\ref{app_proof}. After resumming the ``coeff'' associated with the same combination of transmutation operators $\hat{\partial}_{\sigma_1}\hat{\partial}_{\sigma_2}\cdots\hat{\partial}_{\sigma_l}$, we finally obtain the closed inverse operators. Though, the resummation becomes extremely complicated at higher multiplicity, it indicates that the inverse operator $\hco^{-1}$ takes the same form as~\eqref{eq_def_hco}
\begin{equation}
    \hco^{-1}_{[n-1]}=\left(\sum_{\tau\in [n{-}1]}\prod_{\sigma\in\tau} \tilde{c}_\sigma \hp_\sigma\right)\,.
    \label{eq_def_hcoinverse}
\end{equation}
For example, we could write an ansatz for $\hco_{\{1,2,3\}}^{-1}$ as
\begin{equation}
    \hco_{\{1,2,3\}}^{-1}=1+\tc_{12}\hp_{12}+\tc_{23}\hp_{23}+\tc_{13}\hp_{13}+\tc_{123}\hp_{123}\,,
\end{equation}
then, we have
\begin{equation}
\begin{aligned}
    \hco_{\{1,2,3\}}^{-1}\hco_{\{1,2,3\}}=&1+[(1-\alpha's_{123})\tc_{123}+2\alpha'(\tr_{123}+\\
    &\tc_{12}V_{231}+\tc_{23}V_{312}+\tc_{31}V_{123})]\hp_{123}+\\
    &\sum_{1\leq i<j\leq 3}\left[(1-\alpha's_{ij})\tc_{ij}+2\alpha'\tr_{ij}\right]\hp_{ij}\,.
\end{aligned}
\end{equation}
Make all the coefficients of each transmutation operator to be zero, we can solve out that
\begin{equation}
\left\{
    \begin{aligned}
        &\tc_{12}=t_{12}\tr_{12}\,,~\text{similar for } \tc_{23} \text{ and } \tc_{13}\,,\\[5pt]
        &\tc_{123}=t_{123}(\tr_{123}+\tc_{12}V_{231}+\tc_{23}V_{312}+\tc_{31}V_{123})\,,
    \end{aligned}
\right.
\label{eq_c12c123}
\end{equation}
where we defined 
\begin{equation}
    t_{\rho}:=(-2\alpha')/(1-\alpha's_{\rho})\,,\, \text{with } s_{\rho}=\bigg(\sum_{i\in\rho} p_{i}\bigg)^2\,.
\end{equation}
For general case, we can establish a general formula of the equations for $\tc_\rho$ and solve itself. In Appendix~\ref{app_proof}, we present the local transmutation argument behind the reduction~\eqref{eq_form1}, explain the Kleiss--Kuijf reorganization of the coefficients, and give a detailed five-point illustration. We also find a recursive definition of $c_\rho$ which satisfy $\tc_{\rho}=c_{\rho_1\rho_2[\rho_3[...[\rho_{m-1}\rho_m]]]}$. For $2\leq|\rho|\leq3$, $c_\rho=\tc_\rho$ is defined in~\eqref{eq_c12c123}. For general $\rho$ with $|\rho|>2$, we have
\begin{equation}
    c_{\rho}:=t_{\rho}\left(\tr_\rho+\sum_{i=1}^{\lfloor|\rho|/2\rfloor}\sum_{\substack{1<|\sigma_i|<|\rho|\,,\\(\tau_1\sigma_1\ldots \tau_i\sigma_i)\subseteq \langle\langle\rho\rangle\rangle_i}}V_{\tau_1}\tc_{\sigma_1}\cdots V_{\tau_i}\tc_{\sigma_i}\right)\,,
    \label{eq_def_crho}
\end{equation}
where $\tau_i$ and $\sigma_i$ are contiguous subsequence of the cycle $\rho$, and $\langle\langle\rho\rangle\rangle_i$ is defined as the collection of cycles that make $2i$ of elements in $\rho$ repeat once, for example, $\langle\langle123\rangle\rangle_1=\{(11223),(12233),(11233)\}$. 
In Fig.\ref{fig_crho}, we give a graphical illustration for $c_{12\ldots m}$, and a example for $m{=}4$ is demonstrated in appendix~\ref{app_proof}.

\subsection{Corollaries: $\DF$ gauge theory family}
The inverse operator representation naturally extends from the bosonic coefficient $A^{\DF{+}{\rm YM}}$ to the broader $\DF$ gauge-theory family shown in Fig.~\ref{fig_gauge_family}, whose role in double-copy constructions follows the standard color--kinematics framework~\cite{Bern:2008qj,Bern:2010ue,Bern:2019prr,Kawai:1985xq}. The relevant Lagrangians are reviewed in Appendix~\ref{app_df_family}, following
Refs.~\cite{Johansson:2017srf,Azevedo:2017lkz,Azevedo:2018dgo}. In the main text, the essential point is that the family can be reached from $A^{\DF{+}{\rm YM}}$ by two operations: transmuting selected gluons into scalars~\cite{Cheung:2017ems}, and taking the undeformed $\DF$ limit $\alpha'{\to}\infty$.

For a scalar trace $\rho$, the transmutation relation that maps YM to YMS amplitudes similarly gives
\begin{equation}
    \begin{aligned}
        &A^{\DF{+}{\rm YM}+\phi^3}_{\rho|\mathbb{I}}
        =\hp_{\rho}A^{\DF{+}{\rm YM}}_{\mathbb{I}}
        =\hp_\rho\hco^{-1}_{\{1,\ldots,n\}\setminus\{\rho_i\}}A^{\rm YM}_{\mathbb{I}}\\
        &=\hp_\rho\hco^{-1}_{\{\bar{\rho}\}}A^{\rm YM}_{\mathbb{I}}
        =\hco^{-1}_{\{\bar{\rho}\}}A^{\rm YMS}_{\rho|\mathbb{I}} \, .
    \end{aligned}
\end{equation}
Here $\rho_i$ is an arbitrary element of $\rho$, 
$\{\rho\}$ denotes the set of elements appearing in $\rho$, 
and $\{\bar{\rho}\}$ is the complement of $\{\rho\}$, 
\textit{i.e.}, $\{\rho\}\cup\{\bar{\rho}\}{=}[n]$. 
In the third equality, we used the multilinearity in the polarization vectors, so that $\hco^{-1}$ only acts on the gluons not included in the scalar trace.

The pure $\DF{+}\phi^3$ amplitudes are obtained by taking the undeformed $\DF$ limit of the scalar extension,
\begin{equation}
    A^{\DF{+}\phi^3}_{\rho|\mathbb{I}}
    =\lim_{\alpha'\to\infty}A^{\DF{+}{\rm YM}+\phi^3}_{\rho|\mathbb{I}}
    =\hat {\cal P}_{\{\bar\rho\}}A^{\rm YMS}_{\rho|\mathbb{I}} \, .
\end{equation}
Here $\hat {\cal P}_{\{\rho\}}=\lim_{\alpha'\to\infty}\hco^{-1}_{\{\rho\}}$.
Note that all $\alpha'$ dependence appears only in the variables $t_\rho$, 
which enter the definition of the $c$ variables.
Since $\lim_{\alpha'\to\infty}t_\rho=2/s_\rho$, we obtain
\begin{equation}
    \hat {\cal P}_{\{\rho\}}
    =\hco^{-1}_{\{\rho\}}|_{c\to d}\,,
    \quad
    d_\sigma:=\lim_{\alpha'\to\infty}c_\sigma
    =c_\sigma|_{t\to 2/s}\, .
    \label{eq_def_hcp}
\end{equation}

Similarly, pure-gluon $\DF$ amplitudes follow from the same limit of the massive deformation,
\begin{equation}
    A^{\DF}_{\mathbb{I}}
    =\lim_{\alpha'\to\infty}(-\alpha')^{-1}A^{\DF{+}{\rm YM}}_{\mathbb{I}}
    =\hat {\cal Q}_{\{\mathbb{I}\}}A^{\rm YM}_{\mathbb{I}} \, ,
\end{equation}
where $\hat {\cal Q}_{\{\mathbb{I}\}}
{=}\lim_{\alpha'\to\infty}(-\alpha')^{-1}\hco^{-1}_{\{\mathbb{I}\}}$.
For $\DF$ amplitudes, we encounter the special kinematic situation
$t_\rho{=}{-}2\alpha'$ when $|\rho|{=}n{-}1$ in $n$-point massless scattering.
As a consequence, all $c_\sigma$ with $|\sigma|{<}(n{-}1)$ drop out in the
$\alpha'{\to}\infty$ limit.
Hence, we have
\begin{equation}
    \hat{\mathcal{Q}}_{\{\mathbb{I}\}}
    =\hco^{-1}_{\{\mathbb{I}\}}|_{c\to e}\,,
    \quad
    e_\rho:=\delta_{|\rho|,n-1}s_\rho d_\rho\, ,
    \label{eq_def_hcq}
\end{equation}
where the factor $s_\rho$ cancels the simple pole $1/s_\rho$ in $d_\rho$.

The resulting $\DF$ amplitudes admit a more compact closed form,
\begin{equation}
    A^{\DF}_{\mathbb{I}}
    =\sum_{\sigma\in S_{n-3}}
    \tilde{e}_{1\sigma(2\ldots(n-2))(n-1)}
    A^{\rm YMS}_{1\sigma(2\ldots(n-2))(n-1)|\mathbb{I}} \, .
\end{equation}
Notice that $A^{\rm YMS}_{12\ldots(n-1)|\mathbb{I}}$ can be expanded as
\begin{equation}
   A^{\rm YMS}_{12\ldots(n-1)|\mathbb{I}}
   =\sum_{i=1}^{n-2}
   \bigg(\epsilon_n\cdot\sum_{j=1}^{i}k_j\bigg)
   m(1\ldots i\,n\ldots(n-1)|\mathbb{I}) \, ,
\end{equation}
where $m(\rho|\sigma)$ is the bi-adjoint $\phi^3$ amplitude characterized by the two orderings $\rho$ and $\sigma$~\cite{Cachazo:2013iea,Cachazo:2014xea}.
After straightforward algebraic manipulations, we obtain the crossing Bern-Carrasco-Johansson (BCJ) numerator~\cite{Bern:2008qj,Johansson:2017srf,Edison:2020ehu} for $\DF$ amplitudes,
\begin{equation}
    N^{\DF}_{12\ldots n}
    =\tilde{e}_{12\ldots(n-1)}(\epsilon_n\cdot k_{n-1})\, .
    \label{eq_df2_bcjnum}
\end{equation}
This BCJ numerator manifests the gauge invariance of $(n{-}1)$ gluons.
A highly non-trivial observation is that all $\epsilon\cdot\epsilon$ structures
present in the $d$~\eqref{eq_def_hcp} or $e$~\eqref{eq_def_hcq} variables cancel identically, which is consistent with
the defining properties of $\DF$ and $\DF{+}\phi^3$ amplitudes.

We close this section by comparing our representation with two existing approaches.
First, Ref.~\cite{Chen:2024gkj} constructed crossing-symmetric BCJ numerators for $\DF{+}{\rm YM}$ using kinematic Hopf algebra and the heavy-mass factorization limit. In that language, our coefficient $\tc_\rho$ coincides with $W'(\rho)$ defined in Eq.~(28) of~\cite{Chen:2024gkj}. Thus the inverse operator gives a gauge-invariant amplitude-level origin of the same coefficient data, and one solution to the crossing BCJ numerators of $\DF{+}{\rm YM}$ can be generated directly from YM numerators~\cite{Chen:2024gkj,Edison:2020ehu} as
\begin{equation}
N^{\DF{+}{\rm YM}}=\hco^{-1} N^{\rm YM}\,.
\label{eq_df2ym_bcj}
\end{equation}
A nontrivial observation is that $\hco^{-1} N^{\mathrm{YM}}$ is independent of the adjoint orderings of the transmutation operators appearing in $\hco^{-1}$, despite the fact that $N^{\mathrm{YM}}$ itself is not gauge invariant~\footnote{We have checked~\eqref{eq_df2ym_bcj} up to seven points.}.

Second, our coefficient $c_\rho$ agrees with the coefficient $T_\rho$ defined in Eq.~(16) of~\cite{He:2018pol}, up to an overall factor of $-2\alpha$. This identifies the tachyon-pole coefficients appearing in our transmutation formula for $\DF{+}{\rm YM}+\phi^3$ with the coefficients obtained from the IBP reduction of bosonic-string integrands~\cite{He:2018pol,He:2019drm}. The explicit $\DF{+}{\rm YM}$ amplitudes in~\cite{He:2018pol,He:2019drm} are written in a different basis, and we have verified the equivalence through seven points.

\section*{Conclusion and Outlook}

In this letter, we show that within a restricted EFT framework the bosonic string effective amplitudes are uniquely fixed. This uniqueness leads to nontrivial relations to YMS amplitudes and directly unifies several classes of gauge theories, yielding new results such as compact formulas for amplitudes and even for the BCJ numerators of the $\DF$ theory.

Our result~\eqref{eq-results}, obtained from Eqs.\eqref{eq-inverse} and\eqref{eq_def_hcoinverse}, uncovers the tachyon-pole structure encoded in the coefficients $\tilde{c}_\sigma$ of tree-level bosonic string amplitudes. Remarkably, this structure is expressed in a form that is directly organized by the corresponding field-theory amplitudes $A^{\rm YMS}_{\sigma|\mathbb{I}}$
\begin{equation}\label{eq-results}
A_n^{\DF{+}{\rm YM}}=\left(\sum_{\tau\in [n{-}1]}\prod_{\sigma\in\tau} \tilde{c}_\sigma A^{\rm YMS}_{\sigma|\mathbb{I}}\right)\,.
\end{equation}
While our current work focuses on tree-level open strings, the construction can be naturally extended to closed strings via the double copy, preserving both the uniqueness and analytic structures. 

It would be interesting to further explore the implications for mixed disk string amplitudes~\cite{Stieberger:2009hq,Stieberger:2015vya}, particularly in the framework of stringy form factors~\cite{Cao:2025hio}. This question is particularly natural in light of recent progress on one-loop open-string double-copy structures and one-loop KLT-type relations~\cite{Mafra:2017ioj,Edison:2021ebi,Stieberger:2022lss,Stieberger:2023nol,Cao:2024olg,Cao:2025ygu}. It would be interesting to determine whether an analogue of~\eqref{eq:new eq super} survives at loop level, and whether the tachyon-pole coefficients found here continue to organize the bosonic-string integrand or only emerge after suitable IBP reductions.

\section*{Acknowledgements}
We thank for comprehensive discussions with Song He, Dongyu Yang and Yong Zhang. The work of Q.C. is supported by the Westlake Fellows Program at Westlake University. 
The work of F.Z. is supported in part by the Science Challenge Project (No. TZ2025012), and NSAF No. U2330401.


\bibliographystyle{apsrev4-1}
\bibliography{Refs}

\appendix
\onecolumngrid
\section{Short introduction to the $\DF$ family}\label{app_df_family}
We briefly review the $\DF$ gauge-theory family following Refs.~\cite{Johansson:2017srf,Azevedo:2018dgo}. These theories are useful because their tree amplitudes obey color--kinematics duality and can therefore serve as gauge-theory factors in double-copy constructions. With
\begin{equation}
	F^3:=f^{abc}F^a_{\mu}{}^{\nu}F^b_{\nu}{}^{\rho}F^c_{\rho}{}^{\mu}\,,
\end{equation}
the basic dimension-six theory introduced in~\cite{Johansson:2017srf} is
\begin{equation}
\begin{aligned}
	{\cal L}_{\DF}
	=&\,\frac12(D_\mu F^{a\mu\nu})^2-\frac13 gF^3
	+\frac12(D_\mu\varphi^\alpha)^2
	+\frac12g\,C^{\alpha ab}\varphi^\alpha F^a_{\mu\nu}F^{b\mu\nu}
	+\frac{1}{3!}g\,d^{\alpha\beta\gamma}
	\varphi^\alpha\varphi^\beta\varphi^\gamma .
\end{aligned}
\label{app_lag_df2}
\end{equation}
The scalar $\varphi^\alpha$ transforms in a real representation of the gauge group. The invariant tensors $C^{\alpha ab}$ and $d^{\alpha\beta\gamma}$ are chosen so that the amplitudes admit color--kinematics-dual numerators. Thus the scalar sector is part of the color--kinematics-compatible completion of the higher-derivative gauge theory. Double copying this theory with Yang--Mills gives conformal gravity, and replacing Yang--Mills by supersymmetric Yang--Mills gives the corresponding conformal supergravity amplitudes~\cite{Johansson:2017srf}.

Ref.~\cite{Azevedo:2018dgo} further showed that the gauge-theory building block of bosonic-string amplitudes is a mass deformation of this theory. In our conventions this $\DF{+}{\rm YM}$ theory can be written as
\begin{equation}
	{\cal L}_{\DF+{\rm YM}}
	={\cal L}_{\DF}
	-\frac12m^2(\varphi^\alpha)^2
	-\frac14m^2(F^a_{\mu\nu})^2 .
\label{app_lag_df2ym}
\end{equation}
The mass parameter is identified with the string tension as $m^2=-1/\alpha'$. This makes the tachyon pole of the bosonic string manifest in field-theory language. In the limit $m\to0$ one recovers the undeformed $\DF$ theory, while for large $m$ the higher-derivative states decouple and the amplitudes reduce to Yang--Mills. The inverse-string-tension effective Lagrangian obtained from this mass deformation is known through ${\cal O}(\alpha'^4)$~\cite{Garozzo:2024myw}.

The scalar extensions used for heterotic-string and gauge-gravity sectors are obtained by adding a bi-adjoint scalar $\phi^{aA}$, where $a$ is a gauge index and $A$ is an adjoint index of a second color group with structure constants $\tilde f^{ABC}$:
\begin{equation}
\begin{aligned}
	{\cal L}_{\DF+\phi^3}
	=&\,{\cal L}_{\DF}
	+\frac12(D_\mu\phi^{aA})^2
	+\frac12g\,C^{\alpha ab}\varphi^\alpha\phi^{aA}\phi^{bA}
	+\frac{1}{3!}g\lambda f^{abc}\tilde f^{ABC}
	\phi^{aA}\phi^{bB}\phi^{cC},\\
	{\cal L}_{\DF+{\rm YM}+\phi^3}
	=&\,{\cal L}_{\DF+\phi^3}
	-\frac12m^2(\varphi^\alpha)^2
	-\frac14m^2(F^a_{\mu\nu})^2 .
\end{aligned}
\label{app_lag_df2_family}
\end{equation}
Together with the mass deformation in~\eqref{app_lag_df2ym}, this gives the $\DF{+}{\rm YM}{+}\phi^3$ theory. These scalar theories interpolate between Yang--Mills-scalar amplitudes and their higher-derivative $\DF$ counterparts, and provide the gauge-theory factors that enter the heterotic-string double-copy representation~\cite{Azevedo:2018dgo}.

From this perspective, the results of the main text are not merely compact formulas for a special gauge theory. They provide closed, all-multiplicity expressions for the color--kinematics-compatible building blocks that enter bosonic string, heterotic string, and conformal gravity double copies. The inverse-operator construction also makes the relation between the finite-$\alpha'$ massive deformation and the $\alpha'\to\infty$ $\DF$ limits explicit.

In Table~\ref{table_df2}, we summarize the standard double-copy interpretation of these theories.
\begin{table}[htbp]
	\centering
	\begin{tabular}{c||c}
		$\otimes$&  ($\mathcal{N}=0,1,2,4$) SYM\\[2pt]
		\hline\hline
		$\DF$& Conformal SG\\[2pt]
		$\DF$+YM& Weyl-Einstein SG\\[2pt]
		$\DF$+YM+$\phi^3$& Yang-Mills-Weyl SG\\[2pt]
		$\DF{+}\phi^3$& Yang-Mills-Weyl-Einstein SG
	\end{tabular}
	\captionsetup{justification=raggedright}
	\caption{Various known double-copy constructions for $\DF$ theory and its extensions.}
	\label{table_df2}
\end{table}

\section{Examples for $\DF$ family}\label{app_exp}

In this appendix, we provide several explicit low-point examples of bosonic amplitudes in the $\DF+YM$ and $\DF$ theories, together with their scalar extensions.

\subsection{$\DF+YM$ theory and its scalar extension}

\noindent[3-pt]:
\begin{equation}
    A_{123}^{\DF+\rm YM}=
    A_{123}^{\rm YM}+\tc_{12} A_{12|123}^{\rm YMS}
\end{equation}

\noindent[4-pt]:
\begin{equation}
    A_{1234}^{\DF+\rm YM}=
    A_{1234}^{\rm YM}+\tc_{12} A_{12|1234}^{\rm YMS}+\tc_{13} A_{13|1234}^{\rm YMS}+\tc_{23} A_{23|1234}^{\rm YMS}+\tc_{123} A_{123|1234}^{\rm YMS}
\end{equation}

\begin{equation}
    \begin{aligned}
        A_{34|1234}^{\DF+{\rm YM}+\phi3}=A_{34|1234}^{\rm YMS}+\tc_{12}A_{12,34|1234}^{\rm YMS}
    \end{aligned}
\end{equation}

\noindent[5-pt]:
\begin{equation}
    \begin{aligned}
        A_{12345}^{\DF+\rm YM}&=\,
        A_{12345}^{\rm YM}+\tc_{12} A_{12|12345}^{\rm YMS}+\tc_{13} A_{13|12345}^{\rm YMS}+\tc_{14} A_{14|12345}^{\rm YMS}+\tc_{23} A_{23|12345}^{\rm YMS}+\tc_{24} A_{24|12345}^{\rm YMS}+\tc_{34} A_{34|12345}^{\rm YMS}\\
        &+\tc_{123} A_{123|12345}^{\rm YMS}+\tc_{124} A_{124|12345}^{\rm YMS}+\tc_{134} A_{134|12345}^{\rm YMS}+\tc_{234} A_{234|12345}^{\rm YMS}+\tc_{1234} A_{1234|12345}^{\rm YMS}+\tc_{1324} A_{1324|12345}^{\rm YMS}\\
        &+\tc_{12} \tc_{34} A_{12,34|12345}^{\rm YMS}+\tc_{13} \tc_{24} A_{13,24|12345}^{\rm YMS}+\tc_{14} \tc_{23} A_{14,23|12345}^{\rm YMS}
    \end{aligned}
\end{equation}

\begin{equation}
    A_{45|12345}^{\DF+{\rm YM}+\phi3}=
    A_{45|12345}^{\rm YMS}+\tc_{12} A_{12,45|12345}^{\rm YMS}+\tc_{13} A_{13,45|12345}^{\rm YMS}+\tc_{23} A_{23,45|12345}^{\rm YMS}+\tc_{123} A_{123,45|12345}^{\rm YMS}
\end{equation}

\begin{equation}
    \begin{aligned}
        A_{145|12345}^{\DF+{\rm YM}+\phi3}=A_{145|12345}^{\rm YMS}+\tc_{23}A_{23,145|12345}^{\rm YMS}
    \end{aligned}
\end{equation}
\subsection{$\DF$ theory and its scalar extension}

\noindent[3-pt]:
\begin{equation}
    A_{123}^{\DF}=\tilde{e}_{12} A_{12|123}^{\rm YMS},\quad \tilde{e}_{12}=e_{12}=2\tr_{12}
\end{equation}

\noindent[4-pt]:
\begin{equation}
    A_{1234}^{\DF}=\tilde{e}_{123} A_{123|1234}^{\rm YMS},\quad \tilde{e}_{123}=e_{123}=2\left(\tr_{123}+\frac{\tilde{e}_{12} V_{231}}{s_{12}}+\frac{\tilde{e}_{31} V_{123}}{s_{31}}+\frac{\tilde{e}_{23} V_{312}}{s_{23}}\right)
\end{equation}

\begin{equation}
    A_{12|1234}^{\DF+\phi^3}=A_{12|1234}^{\rm YMS}+\tilde{d}_{34} A_{34,12|1234}^{\rm YMS},\quad \tilde{d}_{34}=d_{34}=\frac{2\tr_{34}}{s_{34}}
\end{equation}

\noindent[5-pt]:
\begin{equation}
\begin{matrix}
    A_{12345}^{\DF}=\tilde{e}_{1234} A_{1234|12345}^{\rm YMS}+\tilde{e}_{1324} A_{1324|12345}^{\rm YMS},\quad \tilde{e}_{1234}=e_{1234}-e_{1243}\,,\\[8pt]
    e_{1234}=2\left[\tr_{1234}+\left(\frac{\tilde{e}_{12}V_{2341}}{s_{12}}+\frac{\tilde{e}_{123}V_{341}}{s_{123}}+3\,\text{cyclic.}\right)+\left(\frac{\tilde{e}_{12}V_{23}\tilde{e}_{34}V_{41}}{s_{12}s_{34}}+1\,\text{cyclic.}\right)\right]
\end{matrix}
\end{equation}

\begin{equation}
\begin{matrix}
    A_{45|12345}^{\DF+\phi^3}=
    A_{45|12345}^{\rm YMS}+\tilde{d}_{12} A_{12,45|12345}^{\rm YMS}+\tilde{d}_{13} A_{13,45|12345}^{\rm YMS}+\tilde{d}_{23} A_{23,45|12345}^{\rm YMS}+\tilde{d}_{123} A_{123,45|12345}^{\rm YMS}\,,\\[8pt]
    \tilde{d}_{123}=d_{123}=\frac{2}{s_{123}}\left(\tr_{123}+\tilde{d}_{12} V_{231}+\tilde{d}_{31} V_{123}+\tilde{d}_{23} V_{312}\right)
\end{matrix}
\end{equation}

\begin{equation}
    \begin{aligned}
        A_{145|12345}^{\DF+\phi3}=A_{145|12345}^{\rm YMS}+\tilde{d}_{23}A_{23,145|12345}^{\rm YMS}
    \end{aligned}
\end{equation}

\section{Transmutation operators}\label{app_transop}
\begin{figure}[htbp]
    \centering
    \includegraphics{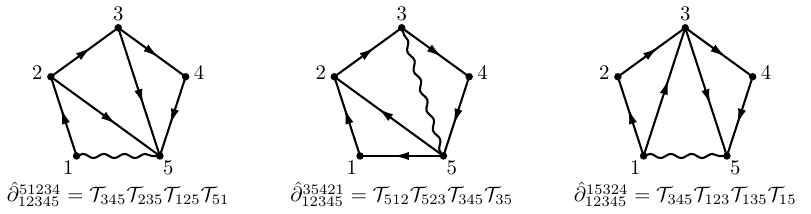}
    \captionsetup{justification=raggedright,singlelinecheck=false}
    \caption{Examples and graphical illustrations of $\hp_{12345}$ with different adjoint orderings. The wave line represents the trace operator $\cT_{ij}$, while the folded line with arrows represents the insertion operator $\cT_{lmn}$.}
    \label{fig_trans_op}
\end{figure}

In this appendix, we present several properties of the transmutation operators~\eqref{eq_def_transop}. Some examples with different adjoint orderings have already been given in~\eqref{eq_exp_hp}.  In general, for a transmutation operator $\hp_{\rho}^{\sigma}$ with $|\rho|=m$, it can be represented graphically by a triangulated $m$-gon, as shown in Fig.~\ref{fig_trans_op}.

As illustrated in Fig.~\ref{fig_trans_op}, each adjoint ordering corresponds to a triangulation with a distinguished edge or chord represented by a wave line. 
Consequently, for a transmutation operator $\hp_{\rho}$ with $|\rho|=m$, there are $(2m-3)C_{m-2}$ distinct combinations of $\cT$ operators, where $C_{m-2}$ denotes the $(m{-}2)$-th Catalan number. 
If the adjoint ordering is fixed to be $\tau$, the set $\{\hp_{\sigma(\rho)}^{\tau}\mid\sigma\in S_m\}$ can be reduced to a Kleiss--Kuijf basis
$\{\hp_{\rho_1\sigma(\rho_2\ldots\rho_{m-1})\rho_m}^{\tau}\mid\sigma\in S_{m-2}\}$,
so that only $(m-2)!$ operators are independent.

The proof is straightforward. 
First, the dihedral symmetry is manifest from the graphical representation in Fig.~\ref{fig_trans_op}. 
Next, consider the operator $\hp_{a_1a_2\ldots a_m}^{\tau}$ and multiply it by the vanishing combination
\begin{equation}
    \begin{aligned}
        0&=\left(\cT_{a_1 a_0 a_2}+\cT_{a_2 a_0 a_3}+\cdots+\cT_{a_m a_0 a_1}\right)\hp_{a_1 a_2 \ldots a_m}^{\tau}\\
        &=\hp_{a_1a_0a_2\ldots a_m}^{\tau a_0}
        +\hp_{a_1a_2a_0\ldots a_m}^{\tau a_0}
        +\cdots
        +\hp_{a_1a_2\ldots a_ma_0}^{\tau a_0}\,,
    \end{aligned}
\end{equation}
which allows one to reduce the original $m!$ operators to a Kleiss--Kuijf basis.
By a similar argument, one obtains the following useful relation,
\begin{equation}
    \hp^{\tau_i\sigma_i\eta}_{\bra{\tau_i}\rho\ket{\sigma_i}}
    \hp^{\theta}_{\tau_i\beta\sigma_i\gamma}
    =
    \hp^{\theta\eta}_{\tau_i(\rho\shuffle\beta)\sigma_i\gamma}
    :=
    \sum_{\zeta\in\rho\shuffle\beta}
    \hp^{\theta\eta}_{\tau_i\zeta\sigma_i\gamma}\,,
    \label{app_eq_hpKK}
\end{equation}
where $\shuffle$ denotes the shuffle product, $\rho$, $\beta$, and $\gamma$ are sequences, and $\eta$ and $\theta$ label adjoint orderings. 
Here $\hp^{\tau_i\sigma_i\eta}_{\bra{\tau_i}\rho\ket{\sigma_i}}$ represents the product of insertion operators appearing in $\hp^{\tau_i\sigma_i\eta}_{\tau_i\rho\sigma_i}$, equivalently defined by
\begin{equation}
    \hp^{\tau_i\sigma_i\eta}_{\bra{\tau_i}\rho\ket{\sigma_i}}
    \cT_{\tau_i\sigma_i}
    :=
    \hp^{\tau_i\sigma_i\eta}_{\tau_i\rho\sigma_i}\,.
\end{equation}
As an example, $\hp^{15234}_{\bra{1}234\ket{5}}=\cT_{125}\cT_{235}\cT_{345}$. 

\section{The recursion structure of coefficients}\label{app_proof}
In this appendix we describe the algebraic mechanism behind the recursion structure of the coefficients $\tc_\rho$. We begin with the linear transformations of $\tc_\rho$ induced by Kleiss--Kuijf relations of transmutation operators. We then combine these transformations with the local action of transmutation operators on $\tr$ and $V$ structures. This proves the reduction used in~\eqref{eq_form1} and shows how the cyclic blocks entering~\eqref{eq_def_crho} arise. We stop at this point, rather than attempting a complete all-multiplicity resummation proof, and close with an explicit five-point illustration.

We use $\rho_i$ for a single label, $\pi^i$ for a possibly empty ordered sequence, and $\pi^i_j$ for the $j$-th element of $\pi^i$. When $j>0$, $\pi^i_{-j}$ denotes the $j$-th element counted from the end of $\pi^i$.

\subsection{Kleiss--Kuijf transformations of the coefficients}

The ansatz for $\hco^{-1}$ is written in terms of transmutation operators. Since these operators obey Kleiss--Kuijf relations, the ordinary coefficients $\tc_\rho$ must transform dually under a change of KK basis. This is a linear basis-independence property of $\tc_\rho$; it should not be confused with the nonlinear recursion formula~\eqref{eq_def_crho}.

At five points, for example, basis independence requires
\begin{equation}
\begin{aligned}
    \tc_{1234}\hp_{1234}+\tc_{1324}\hp_{1324}&=\tc_{1243}\hp_{1243}+\tc_{1423}\hp_{1423}\,,\\
    \tc_{1234}\hp_{1234}+\tc_{1324}\hp_{1324}&=\tc_{1243}(-\hp_{1234}-\hp_{1324})+\tc_{1423}\hp_{1324}\,.
\end{aligned}
\end{equation}
Thus $\tc_{1234}=-\tc_{1243}$ and $\tc_{1324}=-\tc_{1[24]3}$. More generally,
\begin{equation}
    \begin{aligned}
        \sum_{\rho^1,\rho^2}\tc_{\sigma_1\rho^1\sigma_2\rho^2\sigma_3}\hp_{\sigma_1\rho^1\sigma_2\rho^2\sigma_3}&=\sum_{\rho^3,\rho^4}\tc_{\sigma_1\rho^3\sigma_3\rho^4\sigma_2}\hp_{\sigma_1\rho^3\sigma_3\rho^4\sigma_2}\,,\\
        \sum_{\rho^1,\rho^2}(-1)^{1+|\rho^2|}\tc_{\sigma_1\rho^1\sigma_2\rho^2\sigma_3}\hp_{\sigma_1(\rho^1\shuffle\sigma_3\rho^{2,T})\sigma_2}&=\sum_{\rho^3,\rho^4}\tc_{\sigma_1\rho^3\sigma_3\rho^4\sigma_2}\hp_{\sigma_1\rho^3\sigma_3\rho^4\sigma_2}\,,
    \end{aligned}
\end{equation}
where the sums run over decompositions of the same set of labels on both sides. This gives
\begin{equation}
    \tc_{\sigma_1\rho^3\sigma_3\rho^4\sigma_2}=\sum_{(\theta^1\shuffle\theta^2)\ni\rho^4}(-1)^{1+|\theta^2|}\tc_{\sigma_1\rho^3\theta^1\sigma_2\theta^{2,T}\sigma_3}=-\tc_{\sigma_1\rho^3[[\rho^4\sigma_2]]\sigma_3}\,,
    \label{app_eq_cKK}
\end{equation}
where $[[\rho^i]]$ denotes the right-nested commutator of $\rho^i$. For example,
$\tc_{12345}=-\tc_{1[[345]]2}=-\tc_{1[3[45]]2}=-\tc_{13452}+\tc_{13542}-\tc_{15432}+\tc_{14532}$.

\subsection{Recursive blocks from transmutation}

Let us expand $\hco^{-1}_{[n-1]}\hco_{[n-1]}$ and focus on the terms contributing to a longest single-cycle transmutation operator, for instance $\hp_{12\ldots(n-1)}$. For convenience, we fix the adjoint ordering of each $\hp_\rho$ as $(\rho_{|\rho|}\rho_1\ldots\rho_{|\rho|-1})$.

Since the longest single-cycle transmutation operator contains only one trace operator, all trace operators in the left factor must act on the trace coefficient in the right factor. Thus the relevant local building blocks have the form
\begin{equation}
    \tc_{\{\pi\}}\hp_{\{\pi\}}\left( {\tr}_{\tau}\hp_{\tau}\right)\,,
    \label{eq_proof_form}
\end{equation}
where $\hp_{\{\pi\}}=\hp_{\pi^1}\hp_{\pi^2}\cdots$, and similarly for $\tc_{\{\pi\}}$. The elementary actions needed below are
\begin{equation}
\begin{matrix}
    \cT_{i_1 i_2 }\tr_{\sigma^1 j_1j_2\sigma^2}=\delta_{i_1 j_1}\delta_{i_2 j_2}V_{j_2\sigma^2\sigma^1j_1}\\[5pt]
    \cT_{i_1 i_2 }V_{\sigma^1 j_1j_2\sigma^2}=\delta_{i_1 j_1}\delta_{i_2 j_2}V_{\sigma^1 j_1}V_{j_2\sigma^2}\\[5pt]
    \cT_{i_1 i_2 i_3}V_{j_1j_2\sigma^1 j_3j_4}=\delta_{i_1 j_1}\delta_{i_2j_2}V_{j_2\sigma^1 j_3j_4}+\delta_{i_2 j_3}\delta_{i_3j_4}V_{j_1j_2\sigma^1 j_3}
\end{matrix}
\label{eq_proof_TV}
\end{equation}

Here $\cT_{i_1i_2i_3}$ never acts on objects involving the endpoint polarizations $\epsilon_{i_1}$ or $\epsilon_{i_3}$. Therefore
\begin{equation}
    \hp_{\{\pi\}} \tr_{\tau}\sim (\text{Product of $V$ variables})\times(\text{Product of insertion operators $\cT$})
\end{equation}
and the ordered sequences inside each $V$ can only be split or shortened while preserving their relative order. A concrete five-point example of this action is displayed in the final subsection.

We now spell out the general local pattern. Denote the trace operators in $\hp_{\pi^i}$ by $\cT_{\sigma_i\rho_i}$. A nonzero contribution requires that $\sigma_i,\rho_i\in\pi^i\cap\tau$ and that they are adjacent in the trace $\tr_\tau$. Thus we may write
$\tau=\sigma_1\rho_1\gamma^1\sigma_2\rho_2\gamma^2\cdots$, and obtain
\begin{equation}
    \left(\hp_{\sigma_1\rho_1}\hp_{\sigma_2\rho_2}\cdots\right) \tr_{\sigma_1\rho_1\gamma^1\sigma_2\rho_2\gamma^2\cdots}=V_{\rho_1\gamma^1\sigma_2} V_{\rho_2\gamma^2\sigma_3}\cdots\,.
    \label{eq_proof_step1}
\end{equation}
Next consider the remaining insertion operators in $\hp_{\pi^i}$. With our fixed adjoint ordering, write $\pi^i=(\sigma_i \omega^i \rho_i)$, so that these operators are
$\{\cT_{\sigma_i\omega^i_{1}\rho_i},\cT_{\omega^i_1\omega^i_{2}\rho_i},\cT_{\omega^i_2\omega^i_{3}\rho_i},\ldots\}$.
The first insertion operator can act nontrivially only on the neighboring factors $V_{\rho_i \gamma^i \sigma_{i+1}}$ or $V_{\rho_{i-1} \gamma^{i-1}\sigma_i}$. For the former case,
\begin{equation}
    \cT_{\sigma_i \omega^i_1 \rho_i}V_{\rho_i \gamma^i_1\gamma^i_2\ldots \sigma_{i+1}}
    =-\delta_{\omega^i_1\gamma^i_1}V_{\gamma^i_1\gamma^i_2\ldots \sigma_{i+1}}\,,
\end{equation}
then, for $\cT_{\omega^i_1 \omega^i_2 \rho_i}$, we have
\begin{equation}
    \left(\cT_{\omega^i_1 \omega^i_2 \rho_i}\cT_{\sigma_i \omega^i_1 \rho_i}\right)V_{\rho_i \gamma^i_1\gamma^i_2\ldots \sigma_{i+1}}=-\left(\delta_{\omega^i_2\gamma^i_2}\delta_{\omega^i_1\gamma^i_1}\right)V_{\gamma^i_2\gamma^i_3\ldots \sigma_{i+1}}\,.
\end{equation}
Generally, we have
\begin{equation}
    \left(\cT_{\omega^i_{m-1} \omega^i_m \rho_i}\cdots\cT_{\omega^i_1 \omega^i_2 \rho_i}\cT_{\sigma_i \omega^i_1 \rho_i}\right)V_{\rho_i \gamma^i_1\gamma^i_2\ldots \sigma_{i+1}}=-\left(\delta_{\omega^i_{m}\gamma^i_m}\cdots\delta_{\omega^i_2\gamma^i_2}\delta_{\omega^i_1\gamma^i_1}\right)V_{\gamma^i_m\gamma^i_{m+1}\ldots \sigma_{i+1}}\,.
    \label{eq_proof_TV1}
\end{equation}
Similarly, we have
\begin{equation}
        \left(\cT_{\omega^i_{m-1} \omega^i_m \rho_i}\cdots\cT_{\omega^i_1 \omega^i_2 \rho_i}\cT_{\sigma_i \omega^i_1 \rho_i}\right)V_{\rho_{i-1}\ldots \gamma^{i-1}_{-2}\gamma^{i-1}_{-1} \sigma_{i}}
        =(-1)^m\left(\delta_{\omega^i_{m}\gamma^{i-1}_{-m}}\cdots\delta_{\omega^i_2\gamma^{i-1}_{-2}}\delta_{\omega^i_1\gamma^{i-1}_{-1}}\right)V_{\rho_{i-1}\ldots \gamma^{i-1}_{-m-1}\gamma^{i-1}_{-m}}\,,
    \label{eq_proof_TV2}
\end{equation}
Equations~\eqref{eq_proof_TV1} and~\eqref{eq_proof_TV2} cannot be nonzero at the same time. Focusing on~\eqref{eq_proof_TV1}, with the other cases treated analogously, we find
\begin{equation}
    \hp_{\bra{\sigma_i}\omega^i\ket{\rho_i}}V_{\rho_i \gamma^i \sigma_{i+1}}=-\left(\delta_{\omega^i_x\gamma^i_{x}}\cdots\delta_{\omega^i_2\gamma^i_2}\delta_{\omega^i_1\gamma^i_1}\right)
    V_{\omega^i_{x}\gamma^i_{x+1}\ldots \sigma_{i+1}}\hp_{\bra{\gamma^i_x}\omega^i_{x+1}\cdots\ket{\rho_i}}\,,
    \label{app_eq_hpV1}
\end{equation}
where $x$ is the largest integer such that the right-hand side does not vanish. Using the definition of the insertion-operator product, the remaining operators
$\hp_{\bra{\gamma^i_{x}}\omega^i_{x+1}\cdots\ket{\rho_i}}$
acting on $\hp_{\sigma_1\rho_1\gamma^1\ldots\sigma_i\rho_i\gamma^i\ldots}$ yield a sum of transmutation operators. This proves the reduction stated in~\eqref{eq_form1}.

We now continue the same local computation and record the part relevant for the coefficient recursion. From~\eqref{app_eq_hpV1}, one obtains
\begin{equation}
	\tc_{\{\pi\}}\hp_{\{\pi\}}\tr_{\sigma_1\rho_1\gamma^1\sigma_2\rho_2\gamma^2\cdots}\hp_{\sigma_1\rho_1\gamma^1\sigma_2\rho_2\gamma^2\cdots}
	=\left(\prod_{i}-V_{\gamma^i_{x(i)}\gamma^i_{x(i)+1}\ldots \sigma_{i+1}}\tc_{\sigma_i\omega^i\rho_i}\hp_{\langle\gamma^i_{x(i)}|\omega^i_{x(i)+1}\ldots|\rho_i\rangle}\right)
	\hp_{\sigma_1\rho_1\gamma^1\sigma_2\rho_2\gamma^2\cdots}
	\label{app_eq_final1}
\end{equation}
where the delta functions have been suppressed and $x(i)$ is the largest integer for which the corresponding term is nonzero. To simplify the right-hand side, define
$\eta^i=\gamma^i_{1}\cdots\gamma^i_{x(i)-1}$,
$\theta^i=\gamma^i_{x(i)+1}\cdots\gamma^i_{-1}$,
$\lambda^i=\omega^i_{x(i)+1}\cdots\omega^i_{-1}$, and
$\zeta_i=\gamma^i_{x(i)}$. Then
\begin{equation}
	\left(\prod_{i}-V_{\zeta_i\theta^i\sigma_{i+1}}
	\tc_{\sigma_i\eta^i\zeta_i\lambda^i\rho_i}\hp_{\langle\zeta_i|\lambda^i|\rho_i\rangle}\right)
	\hp_{\sigma_1\rho_1\eta^1\zeta_1\theta^1\cdots}
	\label{app_eq_final1b}
\end{equation}
 
Using~\eqref{app_eq_cKK}, the expression~\eqref{app_eq_final1b} can be reorganized as
 \begin{equation}
     \begin{aligned}
         &\left(\prod_{i}\sum_{(\mu^i\shuffle\nu^i)\ni\lambda^i}(-1)^{|\nu^i|}\tc_{\sigma_i\eta^i\mu^i\rho_i\nu^{i,T}\zeta_i}V_{\zeta_i\theta^i\sigma_{i+1}}\hp_{\bra{\zeta_i}\lambda^i\ket{\rho_i}}\right)\hp_{\sigma_1\rho_1\eta^1\zeta_1\theta^1\sigma_2\rho_2\eta^2\zeta_2\theta^2\,\cdots}\\
         \overset{\eqref{app_eq_hpKK} }{=}&\left(\prod_{i}\sum_{(\mu^i\shuffle\nu^i)\ni\lambda^i}(-1)^{|\mu^i|}\tc_{\sigma_i\eta^i\mu^i\rho_i\nu^{i,T}\zeta_i}V_{\zeta_i\theta^i\sigma_{i+1}}\right)\hp_{\sigma_1\rho_1(\eta^1\shuffle\lambda^{1,T})\zeta_1\theta^1\sigma_2\rho_2(\eta^2\shuffle\lambda^{2,T})\zeta_2\theta^2\,\cdots}
     \end{aligned}\,,
     \label{app_eq_final2}
 \end{equation}
This is the main point needed for the present paper. The coefficient multiplying the remaining transmutation operator is already organized into cyclic blocks of the type represented in Fig.~\ref{fig_crho}: each block contains a $V$ factor and a lower coefficient $\tc$, and the KK sums fix the right-nested combination of the lower coefficient. A complete closed-form proof of the subsequent all-multiplicity resummation would require controlling the remaining overlap among all KK-basis contributions. We do not attempt that final step here.

\subsection{Five-point illustration}

\begin{figure*}[htbp]
    \centering
    \includegraphics{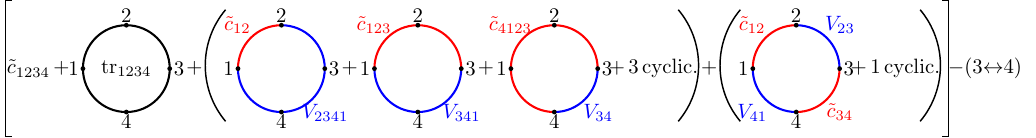}
    \caption{The coefficient of $\hp_{1234}$ in $\hco_{[4]}\hco^{-1}_{[4]}$. The diagrammatic definitions follow those of Fig.~\ref{fig_crho}.}
    \label{fig_tc4_1}
\end{figure*}

At five points, the part of $\hco^{-1}_{[4]}\hco_{[4]}$ relevant for the longest single-cycle sector is
\begin{equation}
    \begin{aligned}
        &\hco^{-1}_{[4]}\hco_{[4]}\\=&(1+\tc_{12}\hp_{12}+\tc_{13}\hp_{13}^{31}+\tc_{23}\hp_{23}+\tc_{123}\hp_{123}+\tc_{124}\hp_{124}+\tc_{134}\hp_{134}+\tc_{234}\hp_{234}+\tc_{1234}\hp_{1234}+\tc_{1324}\hp_{1324}+\tc_{12}\tc_{34}\hp_{12}\hp_{34}\\
        &+\tc_{13}\tc_{24}\hp_{13}\hp_{24}+\tc_{14}\tc_{23}\hp_{14}\hp_{23})\times(1+\tr_{12}\hp_{12}+\tr_{13}\hp_{13}+\tr_{23}\hp_{23}+\tr_{123}\hp_{123}+\tr_{124}\hp_{124}+\tr_{134}\hp_{134}+\tr_{234}\hp_{234}\\
        &+\tr_{1234}\hp_{1234}+\tr_{1324}\hp_{1324}+\tr_{1243}\hp_{1243}+\tr_{12}\tr_{34}\hp_{12}\hp_{34}+\tr_{13}\tr_{24}\hp_{13}\hp_{24}+\tr_{14}\tr_{23}\hp_{14}\hp_{23})\,.
    \end{aligned}
    \label{app_eq_exp_oo}
\end{equation}
For instance, the term $\tc_{123}\hp_{123}^{312}$ gives
\begin{equation}
    \begin{aligned}
        &\tc_{123}\hp_{123}^{312}\left(\tr_{134}\hp_{134}+\tr_{1324}\hp_{1324}+\tr_{1243}\hp_{1243}\right)\\
        =\,&\tc_{123}\cT_{123}\cT_{31}\left(\tr_{134}\hp_{134}+\tr_{1324}\hp_{1324}+\tr_{1243}\hp_{1243}\right)\\
        =\,&\tc_{123}\left(V_{341}\cT_{123}\hp_{134}-V_{241}\hp_{1324}+V_{243}\hp_{1243}\right)\\
        =\,&\tc_{123}\left(V_{341}\hp_{1234}-V_{241}\hp_{1324}+V_{243}\hp_{1243}\right)\\
        = \,&\tc_{123}\left[(V_{341}-V_{243})\hp_{1234}+(-V_{241}-V_{243})\hp_{1324}\right]\,.
    \end{aligned}
\end{equation}
Expanding $\hco_{[4]}\hco^{-1}_{[4]}$ and using the Kleiss--Kuijf relations~\eqref{app_eq_hpKK} and~\eqref{app_eq_cKK}, the coefficient of $\hp_{1234}$ is illustrated in Fig.~\ref{fig_tc4_1}.

\begin{figure*}[htbp]
    \centering
    \includegraphics{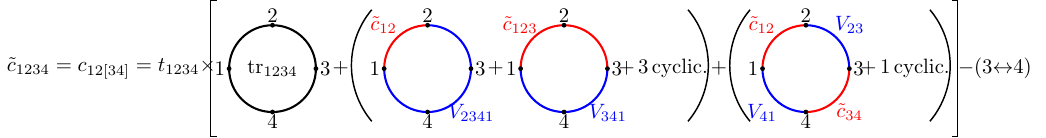}
    \caption{The recursive definition of $\tc_{1234}$.}
    \label{fig_tc4_2}
\end{figure*}

Compared with Fig.~\ref{fig_crho}, Fig.~\ref{fig_tc4_1} still contains red arcs carrying all four nodes. Their sum is precisely the term $s_{1234}\tc_{1234}$ coming from the direct self-contribution of the longest coefficient. Setting the full coefficient of $\hp_{1234}$ to zero then determines $\tc_{1234}$ in terms of lower coefficients and trace structures, as shown in Fig.~\ref{fig_tc4_2}.

\end{document}